\documentstyle[aps]{revtex}
\begin{document}
\baselineskip=25pt
\begin{center}
THE LORENTZ-DIRAC EQUATION  AND THE PHYSICAL MEANING OF THE MAXWELL'S
FIELDS\\
\bigskip
Manoelito Martins de Souza \footnote{
Universidade Federal do Esp\'{\i}rito Santo\\
Departamento de F\'{\i}sica \\
 29060-900  Vit\'oria - ES - Brasil \\
e-mail: manoelit@cce.ufes.br}\\
\end{center}
\medskip
\begin{center}
ABSTRACT
\end{center}
\noindent Classical Electrodynamics is not a consistent theory because of its
field inadequate
behaviour in the vicinity of their sources. Its problems with the electron
equation of motion
and with non-integrable singularity of the electron self field and of its
stress tensor are well
known. These inconsistencies are eliminated if the discrete and localized
(classical photons)
character of the electromagnetic interaction is anticipatively recognized
already in a classical
context. This is possible, in a manifestly covariant way, with a new model of
spacetime
structure, shown in a previous paper $^{1}$, that
 invalidates the Lorentz-Dirac equation. For a point classical electron there
is no field singularity, no causality violation and no conflict
with energy conservation in the electron equation of motion.  The
electromagnetic field must be re-interpreted in terms of average flux of
classical photons. Implications  of a singularity-free formalism to field
theory are discussed.

\section{INTRODUCTION}

This paper concern is about 3 old, but still not solved problems of classical
field theory, more
particularly, of Classical Electrodynamics for a point charge: its field
singularity, the non-
integrable singularities of its energy-momentum tensor, and its bizarre
equation of motion for
the electron (the Lorentz-Dirac equation, LDE). They are the reasons for
Classical
Electrodynamics not being considered an entirely consistent theory; it cannot
simultaneously
handle its fields and their sources defined at a same point. We show in this
paper that these
problems are overcome with the anticipated adoption of some notions proper of
the quantum
theory: the granular, that is, the discrete and localized, character  of a
quantum interaction
(classical photons).\\
This paper is a sequel of reference [1] of, where a proposed geometric vision
of the
causality
constraints in field theory and its possible relevance to the problems of the
LDE are
discussed. The discussion of this geometry is repeated in an amplified way, in
section III. The
following 2 sections are applications of these ideas to the Lienard-Wiechert
solution (LWS) of
Classical Electrodynamics: the natural (in the sense of not ad hoc nor forced)
elimination of
the stress tensor non-integrable singularities, in section II, and of the LDE
problems in
section IV. The relevance of these ideas to Quantum Field Theory, as also to
the question of
the origin and meaning of mass, is developed elsewhere$^{(2,3)}$, but for the
sake of
completeness, a synopsis is presented in section V, after a short qualitative
discussion on the
meaning of the Maxwell-Faraday concept of field.

 The LDE$^{(4)}$ is the greatest paradox of classical field theory as it cannot
simultaneously preserve  both the causality and the energy conservation,
although there
is nothing in
the premisses for its derivation that justify such violations. It is obtained
from  energy-
momentum conservation  in the LWS
\begin{equation}
\label{1} A=\frac{V}{\rho}{\Bigg|}_{{\tau}_{ret}},\;\;\;for\;\;\;\rho>0,
\end{equation}
in the   limit of $\rho\rightarrow0$. V is the electron 4-velocity and $\rho$
is the invariant
distance between a charge and its field. Taking this limit represents an
extrapolation of
conditions not valid for a region where the electron and its electromagnetic
field are
simultaneously defined. This limit passage, as it is argued in $^{(1)}$, is the
weak point in all
demonstrations of the LDE, because it corresponds to the assumption that the
Minkowski
model of spacetime is valid also in this limit. But, the Minkowski spacetime
does not
represent the necessary causality requirements for the description of two
interacting fields at
a same point. This has been pointed$^{(1)}$ as the source of all problems
associated to the
LDE.\\
Many physicists nowadays, as a consequence of the undisputed success of the
Einstein's
Theory of Special Relativity, see a Minkowski manifold as the natural or even
absolute
description of the spacetime structure of the world --- in the absence of
gravity, it is added.
There is nothing special with the Minkowski spacetime. It just represents a
geometric
implementation of a well established principle of physics: the speed of light
is a universal
constant.  It is a geometrization of this physical principle in the same way
that in General
Theory of Relativity the Newtonian concept of gravitational force is replaced
by the notion of
a curved spacetime. For the geometrization of special relativity, time,
considered before as
just an event-ordering parameter, has to be treated as a fourth dimension of
the world. It
must transform among inertial observers, in exactly the same way that the other
3 bona fide
space dimensions do. These two examples show the importance of geometrization
as a  tool
in theoretical physics: transferring a mandatory physical requirement to the
structure of a
background geometry is a way of assuring its automatic implementation.\\
The postulate of a constant speed of light is connected also to the important
concept of
relativistic causality: no physical object can move faster than light in a
vacuum. It is
geometrically implemented through the lightcone structure, seen as a
restriction of access to
regions of spacetime for physical objects. The importance of relativistic
causality to modern
physics has raised it to the category of a principle. But it is exactly this
principle that is
violated by solutions of the Lorentz-Dirac equation$^{(5)}$. This is a
paradoxical situation
because the Minkowski spacetime is one of the premisses always assumed in its
derivations.
The strategy adopted in reference$^{(1)}$ was to create a model that represents
the
geometrization of the causality implementation found in the LWS, which is also
one of the
premisses in the derivation of the LDE.  In section II we present a commented
review of the
content of$^{(1)}$. The geometrization of the causality constraints that govern
the
propagation of physical objects, transforms these constraints on restrictions
of access to
regions of the spacetime, with possible change on its metric structure.
  In section III the Maxwell formalism for the LWS in this new geometry (with
preservation of
causality) is presented. It is shown then that, irrespective the spacetime
metric structure, the
non-integrable singularity of the stress tensor disappears together with the
terms that would
generate, in the usual treatment, the Schott term and the Teitelboim's
bound-momentum.
Previous attempts on taming these singularities have relied on modifications of
the Maxwell's
theory with ad hoc addition of extra- terms (see for example the
reviews$^{(9-11)}$) to the
field stress tensor on the electron worldline; it is particularly interesting
that, as we will show
here, instead of adding anything we should actually not drop out some null
terms. Is their
contribution (not null, in the limit) that avoid the infinities.\\
The same problem happens in the derivations of the electron equation of motion:
they are
done with incomplete field expressions that do not contain these terms that are
null only off
the particle worldline. The Schott term in the LDE is a consequence of this; it
does not
appear in the equation when the full field expression is correctly used.\\
The electron equation of motion,
\begin{equation}
\label{anticipation}
m\hbox{\Large a}^{\mu}-<\frac{1}{4}\;{\hbox{\Large a}}^{2}\hbox{\Large
a}^{\mu}>=F^{\mu}_{ext}-<\frac{2{\hbox{\Large a}}^{2}}{3}V^{\mu}>,
\end{equation}
in this geometrical settings, is derived in section IV.
 The absence of the problematic Schott term assures that there will be no
causality violation,
and $<\frac{2}{3}{\hbox{\Large a}}^{2}V^{\mu}>,$ where the brackets represent a
mean or
average, in the place of $\frac{2}{3}{\hbox{\Large a}}^{2}V^{\mu}$, guarantees
the
conservation laws, as explained at the end of section IV. The new term
$<\frac{1}{4}\;{\hbox{\Large a}}^{2}\hbox{\Large a}^{\mu}>$ is
associated to the electron spacetime curvature. The existence or not of this
curvature term
must be determined by experimental means and evaluated by its theoretical
implications.\\
The recognition that the contribution from the electron self-field to its
equation of motion is
realized through average quantities requires a re-evaluation of the physical
meaning of the
Faraday-Maxwell concept of field, as qualitatively discussed in section V. In
this new
geometrical context, it must be seen as an average flux of ``classical photons"
around the
charge. Then, its singularity loses all of its physical content, because it
becomes just a
consequence of working with this average field in the place of the fundamental
one. This new
geometrical structure of spacetime generates a formalism that is free of
singularities; it is
being more properly discussed elsewhere$^{(2)}$, but for the sake of
completeness we
include a synopsis of its main results as well as the fresh and good results of
the work$^{(6-
8)}$  with students (their M.Sc. theses, all in Portuguese), at the end of
section V. The paper
ends with an appendix showing some calculations that have been omitted in
section IV.

\section{GEOMETRY OF CAUSALITY}
Causality is implemented in the LWS through the constraint
\begin{equation}
\label{lcone}
\Delta x.\eta.\Delta x:=\Delta x^{\mu}\eta_{\mu\nu}\Delta
x^{\nu}=0,\;\;\;\;\Delta x^{0}>0,
\end{equation}
where $\Delta x=x-z(\tau)$,  $z(\tau)$ describes the electron world-line,
parameterized by its
propertime $\tau$, and  x is the point where the electromagnetic field (the
photon) is
observed.\\
If $z(\tau_{ret})$ is the solution to (\ref{lcone}), the geometrical meaning of
$\Delta
x.\eta.\Delta x=0$ is of a light-cone with vertex at $z(\tau_{ret})$, and
corresponds to the
requirement that the electromagnetic field emitted at $z(\tau_{ret})$ remain on
its light-cone.
Its extension to the electromagnetic field at any spacetime point requires that
$\Delta
x.\eta.dx=0$ or
\begin{equation}
\label{dlcone}
d\tau+K.\eta.dx=0,
\end{equation}
where
\begin{equation}
\label{K}
K^{\mu}:=\frac{\Delta x^\mu}{\rho}=\frac{\Delta x^\mu}{\Delta\tau},
\end{equation}
is a null 4-vector, $K^{2}=0$, and represents a light-cone generator;
$\rho:=-V.\eta.\Delta x$.
We want to underline here the well known important role of the light-cone
generator, K, in the
propagation of the electromagnetic field. $d\tau$ in  (\ref{dlcone}) is the
change in the
electron proper-time, which is then connected to the required change in x, the
point where
the electromagnetic wave front is being observed. If $d\tau=0$ ({\it no change}
in the
electron position), then $K.dx=0$ and $dx.dx=0$; the wave front {\it moved} a
distance dx
along the light-cone generator $K^{\mu}$. If $d\tau\neq0$, then $K.dx\neq0$ and
$dx.dx\neq0$, which then means that the field emitted at $z(\tau+d\tau)$ must
be observed
not at x but at $x+dx$, which  belongs to the light-cone of
$z(\tau_{ret}+d\tau)$, and not to
the light-cone of $z(\tau_{ret}).$ Then,  $d\tau\neq0$
\linebreak  implies, in (\ref{dlcone}), that dx represents the difference in
the positions of two
distinct wave fronts (photons), each one on a different lightcone, with
vertices at, respectively
z($\tau$) and $z(\tau+d\tau)$. A clear vision of this is important for
understanding the
pictures of the subsequent problems. \\
The restriction (\ref{dlcone}) can be absorbed in the definition of a
directional derivative
\begin{equation}
\label{convective}
 \nabla_{\mu}\equiv  \partial_{\mu}-K_{\mu}\partial_{\tau},
\end{equation}
which is useful in calculations related to (\ref{1}),  because $\tau$ can be
treated then as
independent of x:$$\frac{\partial}{\partial
x^{\mu}}A(x,\tau){\Big|}_{\tau_{ret}}=(\partial_{\mu}-
K_{\mu}\partial_{\tau})A(x,\tau).$$ For the electromagnetic field $\tau$ is
always $\tau_{ret}$.
This is all the causality content in the LWS for the photon, but it is
important to remark that it
does not apply to the electron. An extension$^{(1)}$ of
(\ref{lcone}-\ref{convective}) for the
electron, requires  their replacement, respectively, by
\begin{equation}
\label{electron}
-(\Delta\tau)^{2}=\Delta z.\eta.\Delta z,
\end{equation}
\begin{equation}
\label{delectron}
d\tau+V.\eta.dz=0,
\end{equation}
and
\begin{equation}
\label{econvective}
 \nabla_{\mu}\equiv  \partial_{\mu}-V_{\mu}\partial_{\tau},
\end{equation}
The differences among (\ref{lcone}-\ref{convective}) and
(\ref{electron}-\ref{econvective}) are
an expression of the differences in the causality requirements for massless and
for massive
fields. Observe that there is a dichotomy in the treatment given to the
electron and to its
electromagnetic field. In the set (\ref{lcone}-\ref{convective}) of causality
restrictions imposed
on the  propagation of the electromagnetic field, $\tau$ is the electron
propertime; so it
explicitly depends  on the electron conditions. But the same does not occur for
the
propagation of the electron: (\ref{electron}-\ref{econvective}) does not depend
on any
parameter of the electromagnetic field. This biased description of these two
physical objects
(the electron and its electromagnetic field) does not correspond to what is
directly observed
in nature, where both are equally elementary. The production of photons by an
electron is so
fundamental as the production of electron-positron pairs by a photon. It is
just a question of
charge, 4-momentum and angular momentum conservation. This seems to be already
a clue
that this classical formalism cannot produce a correct description of nature,
not just for being
classical, but for its biased treatment of two equally fundamental physical
objects.\\
Returning to the limit passage of $\rho\rightarrow0,$ in the usual derivations
of the equation
of motion  (LDE) of a classical electron, which involves its four-momentum
conservation, we
can see from
(\ref{convective}), that the energy-momentum tensor obtained from the LWS has
an explicit
dependence on K, which dictates the propagation of the photon, not of the
electron. This
should not be a problem because$^{(1)}$
\begin{equation}
\label{limit}
\lim_{\rho\to0}{K^{\mu}}=V^{\mu}
\end{equation}
and so, in this limit
(\ref{lcone}-\ref{dlcone})$\rightarrow$(\ref{electron}-\ref{econvective}).
Nonetheless, it can still be a problem because the metric structure of
spacetime is wrongly
supposed to be kept Minkowskian in this limit. In order to show why this may
not be correct, it
is necessary first  to (geometrize) incorporate (\ref{lcone}-\ref{econvective})
into a single
background geometrical structure of spacetime.\\
Consider all the physical objects (electrons, electromagnetic fields, etc)
immersed in a  flat 5-
dimensional space, $R_{5}\equiv R_{4}\otimes R_{1}$, whose line elements are
defined by
\begin{equation}
\label{efive}
(\Delta S_{5})^{2}=\Delta x^{M}\eta_{\mbox{\tiny{MN}}}\Delta x^{N}= (\Delta
S_{4})^{2}-(\Delta
x^{5})^{2}=\Delta x.\eta.\Delta x-(\Delta x^{5})^{2},
\end{equation}
where $M,N=1\;to\;5$. Immersed in this larger space, every physical object is
restricted to a
4-dimensional submanifold, its SPACETIME, by
\begin{equation}
\label{gconst}
-(\Delta x^{5})^{2}=\Delta x.\eta.\Delta x.
\end{equation}
In other words, the CHANGES of $x^{5}$ of a physical object is identified with
the CHANGES
of its very propertime, $\Delta x^{5}=\Delta\tau$.  For a physical object,
$(\Delta S_{5})^{2}=-
2(\Delta\tau)^{2},$  always. This identification of $\Delta x^{5}$ of a
physical object with the
variation of its proper-time, $\Delta \tau$, represents the geometrization of
causality for
massless ($\Delta \tau=0$) and for massive  ($\Delta \tau\neq0$) physical
objects. So, the
constraints on the propagation of physical objects become restrictions on their
allowed
domain in $R_{5}$, that is, in the definition of their allowed spacetime.\\
The evolution or propagation of physical objects, in this geometric setting, is
restricted by the
differential of (\ref{gconst}), $\Delta\tau d\tau+\Delta x.\eta.dx=0$, or
\begin{equation}
\label{generator}
d\tau +f.\eta.dx=0,
\end{equation}
and by its induced directional derivative
\begin{equation}
\label{directional}
 \nabla^{f}_{\mu}:=  \partial_{\mu}-f'_{\mu}\partial\tau,
\end{equation}
where  $f^{\mu}=\frac{\Delta x^{\mu}}{\Delta \tau},$
$f'_{\mu}:=\eta_{\mu\nu}f^{\nu},$ and  f is
a  timelike $4$-vector if $d\tau \neq 0$,
or (extending (\ref{generator}) to include) a light-like $4$-vector if $d\tau =
0$. A light-like f
corresponds  to K of (\ref{dlcone}) while a timelike f stands for V of
(\ref{econvective}).\\
Let us now discuss the geometrical constructions behind the basic equations
(\ref{gconst},
\ref{generator} and \ref{directional}). With a more transparent notation,
(\ref{gconst}) may be
written, as
\begin{equation}
\label{hypercone}
(\Delta t)^{2}=(\Delta\tau)^{2}+(\Delta{\vec{x})}^{2},
\end{equation}
\noindent which defines a 4-dimensional hypercone in the local tangent space of
$R_{5}$. It
is a P-CAUSALITY-CONE, with vertex at a point P, a generalization of the
Minkowski light-
cone.
 A light-cone, the domain of massless physical objects, is an intersection of a
causality-cone
and a 4-dimensional hyperplane  defined by $x^{5}=const$.  The interior of a
light-cone is the
projection of a causality-cone on such a $(x^{5}=const)$-hyperplane. Each
observer
perceives an strictly $(1+3)$-dimensional world and his $\Delta x^{5}$
coincides with the
elapsed time measured on his own clock, as required by special relativity; it
represents his
aging, according to his own clock.\\ Eq. (\ref{generator}) may be written as
$f_{M}dx^{M}=0,$
with $f_{M}:=(f'_{\mu},f_{5}=1).$ For a fixed point P, it defines a family of
4-dimensional
hyperplanes parameterized by $f'_{\mu},$ which is both, normal to the
hyperplane and
tangent to the P-causality-cone (\ref{hypercone}). On the other hand, ${\bar
f}^{M}:=({\bar
f}^{\mu}, 1)$ where ${\bar f}^{\mu}=(f^{0},-f^{i})$ is the common
(intersection) 5-vector,
tangent to the P-causality-cone and to the hyperplane that passes by P and is
orthogonal to
$f_{M}$. Both, the normal, $f_{M}$, and the tangent,  ${\bar f}^{M}$, to the
P-hyperplane, are
tangent vectors to the P-causality-cone, at the point P.

At the vertex of the causality-cone, both, the normal to the hyperplane,
$f'_{\mu},$ and its
tangent on the cone, $f^{\mu}$,  represent two cone-generators, in opposition
to each other.
See the figure. They represent motions (4-velocities) in opposite space
directions but with a
same time direction. So, for a given f, there are two solutions for the
intersection of
(\ref{hypercone}) and (\ref{generator}): generator $f^{\mu}:=(f^{0}, {\vec f})$
and generator
${\bar{f}}^{\mu}:=(f^{0},-{\vec{f}}).$ In section V they will be associated to
the description of
creation and annihilation of particles. \\

f in (\ref{generator}) represents the local causality restriction on the
evolution of a physical
object, and the directional derivative (\ref{directional}) describes its
allowed displacement
along its causality-cone generator f.
Every representation of a physical object should be appended by a label f,
symbolizing the
exhibition of such causal restrictions.\\
Let us introduce the concept of  {\bf CAUSALITY-LINE} in contradistinction to
the Minkowski
worldline, which becomes the projection of the first one on a
($x^{5}=const$)-hypersurface
(that is, on its lightcone, or on its interior). The causality-line of a free
physical object is a
causality-cone generator; an object, under an interaction, may change of
causality-line,
defining then, a new causality-cone vertex. A causality-cone vertex is defined
by the
intersection of 3 generators and corresponds to a pictorial description of a
Lagrange
interaction term, like ${\cal L}_{I}=e{\bar\Psi}_{f}A_{f'}\Psi_{f''},$ where,
in this notation (to be
better explained in section V), an incoming field in the generator f, changes
to another
generator f'', after emitting/absorbing a photon in a generator f'. This
picture of cones and
generators is appropriate to the description of discrete interactions among
point (localized)
objects. We will return to this point in section V.\\ The directional
derivative
$\nabla^{f}_{\mu}$ has then a clear and natural meaning: it is the spacetime
evolution of a
physical object, allowed by its causality restrictions. Let $A_{f}$ and
$B_{f'}$ be objects
whose causality restrictions on their evolution are given, respectively, by f
and f', then
\begin{equation}
\label{chainrule}
\partial_{\mu}\Biggl(A{\bigg|}_{f_{M}.dx^{M}=0}B{\bigg|}_{f'_{M}.dx^{M}=0}\Biggr)\equiv(\nabla_{\mu}^{f}A_{f})B_{f'}+A_{f}(\nabla_{\mu}^{f'}B_{f'}),
\end{equation}
where $\nabla_{\mu}^{f}:=\partial_{\mu}-f_{\mu}\partial_{\tau}$ and
$\nabla_{\mu}^{f'}:=\partial_{\mu}-f'_{\mu}\partial_{\tau}$. In other words,
the label f in a
directional derivative, like in $\nabla^{f}$, is defined by the causal
restrictions of the object on
which it is acting. There is no meaning on an expression like
$\nabla^{f}B_{f'},$ unless $f'=f$.

The equations (\ref{lcone}-\ref{econvective}) have been replaced in this
geometric
description by (\ref{gconst}-\ref{directional}), but now with a very
significative difference:
each of these equations involve parameters of just one physical object. For
example, the
change of $\tau$ in (\ref{gconst}) and in (\ref{generator}) for any given
photon is always null.
The electron and the electromagnetic field, in this respect, are being treated
on a same foot;
each one is defined on its appropriate causality-cone generator. This picture
has deep
physical consequences that are connected to fundamental problems of Quantum
Mechanics
and of Quantum Field Theory; they will be briefly discussed in section V, as
they are outside
the scope of this paper.\\
But it is the metric structure induced  by (\ref{generator}) on the  spacetime
of a physical
object that most clearly exposes the departure from a Minkowski structure. We
have from
(\ref{efive}) and (\ref{generator}), that
$(dS_{5})^{2}=dx.\eta.dx-(f'.dx)^{2}=dx.(\eta-f'f').dx,$
for massive objects, and that $(dS_{5})^{2}=dx.\eta.dx$ for massless objects,
as
$d\tau=f'.dx=0.$ Then, the induced metric is given by
\begin{equation}
\label{metric}
g_{\mu\nu}=\cases{\eta_{\mu\nu}& if $m=0$;\cr
                  \eta_{\mu\nu}-f'_{\mu}f'_{\nu} & if $m\neq0$.\cr}
\end{equation}
The distinct causality requirements of  massive  and  of  massless
fields and particles are,  therefore,  represented  by  immersions
with distinct metric structures. \\
This represents a very strong inference: eq. (\ref{metric})  is saying that
the
change in the propertime of a physical object contributes to the length of its
causality-line.
This contribution is materialized  in a Riemannian manifold with the metric
(\ref{metric}).
Observe that it could be in the other way, since the change in the propertime
of a physical
object represents just its aging, and so why should it be taken into account in
the computing
of its relevant line element? We could, just as well, assume that the
physically relevant line
element is given by the worldline, and not by the causality-line; then, the
above $g_{\mu\nu}$
would be a meaningless concept. This would not be the most esthetics Nature's
choice since it
would spoil the beauty of this geometric picture, but it is indeed a question
to be settled by
experimental verifications. We will return to this point in section IV.\\
While $T^{\mu\nu},_{\nu}=0$ for $\rho>0$ remains valid in this new picture, its
limit when
$\rho\rightarrow 0$ is not as simple as fore-assumed because  it may involve
now  a  local
change of manifolds with different metric structure ($\eta\rightarrow\eta-ff$).
The existence of
two distinct metric structures for a massive and a massless field, which has
not been
considered in the determination of the LDE is just one of the reasons that
invalidates it as the
electron equation of motion. As a matter of fact, we will see in the following
sections, that
regardless the metric structure of the physical spacetime, be it Minkowskian or
Riemannian,
just the understanding of this new geometric picture  is enough to clear
electrodynamics from
some of its inconsistencies and to show that the Lorentz-Dirac equation is not
justified.
However, $g_{\mu\nu}$ and $\eta_{\mu\nu}$ produce distinct measurable effects,
as shown
in (\ref{anticipation}),  that can, in principle, be detected in a laboratory.

\section{THE INTEGRABLE STRESS TENSOR}

We can have a better understanding of the meaning of this new geometry,
studying the
behaviour of the retarded Maxwell tensor, $F^{\mu\nu}_{ret}$, and of its
stress-tensor,
$\Theta^{\mu\nu}_{ret}$, both obtained from the retarded Lienard-Wiechert
solution (\ref{1}),
and both singular at the electron world-line, $\rho=0$, an immediate
consequence of (\ref{1}).
In the standard or Minkowskian approach we write
$\Theta_{ret}=\Theta_{2}+\Theta_{3}+\Theta_{4}$, where the indices indicate the
order of the
respective singularities at $\rho=0$.  $\Theta_{2}$, although singular at
$\rho=0$, is
nonetheless integrable. By that it is meant that $\int d^{4}x\Theta_{2}$
exists$^{(9)}$, while
$\Theta_{3}$ and $\Theta_{4}$ are not integrable; they generate, respectively,
the
problematic Schott term in the LDE and a divergent expression, the electron
bound 4-
momentum$^{12}$. The most update prescription$^{(9,11)}$ is to redefine
$\Theta_{3}$ and
$\Theta_{4}$ at the electron worldline in order to make them integrable, but
without changing
them at $\rho>0$, so to preserve the standard results of Classical
Electrodynamics. This is
possible with the use of distribution theory, but it is always an introduction
of something
strange to the system and in an {\it ad hoc} way. The most unsatisfactory
aspect of this
procedure is that it regularizes the above integral but leaves an unexplained
and unphysical
discontinuity in the flux of 4-momentum from the charge worldline:
$\Theta(\rho=0)\neq\Theta(\rho\sim0).$ We show, in this section, that these
problems have
been solved: no discontinuity and no non-integrable singularity in the electron
self field stress
tensor.\\

For a unified treatment of the electron and of its electromagnetic field we
introduce a
parameter $\chi$:
\begin{equation}
\label{chi}
\chi=\cases{0,& if $\rho>0$;\cr
              1,& if $\rho=0$,\cr}$$
\end{equation}
so that the tensor metric, after (\ref{limit},\ref{metric}), may be written as
\begin{equation}
\label{gc}
g_{\mu\nu}=\eta_{\mu\nu}-\chi V'_{\mu}V'_{\nu},
\end{equation}
with $g^{\mu\nu}=\eta^{\mu\nu}+\frac{\chi}{2}V'_{\mu}V'_{\nu}$ and
$V'_{\mu}:=\eta_{\mu\nu}V^{\nu}$, to be distinguished from
$V_{\mu}=g_{\mu\nu}V^{\nu}=(1+\chi)V'_{\mu}.$\\  $\chi$ has also the double
role of pin
pointing the contribution from the non-Minkowskian geometry on the electron
world-line to its
self-field,  stress tensor, and  equation of motion, as well as of allowing us
to turn off the
effects of the inference (\ref{metric}) by just taking $\chi=0,$ even for
$\rho=0$.\\ The metric
(\ref{gc}) induces a covariant derivative, $D$:
$$D_{\alpha}g_{\mu\nu}:= \nabla_{\alpha}g_{\mu\nu}-
g_{\mu\beta}\Gamma^{\beta}_{\nu\alpha}-g_{\beta\nu}\Gamma^{\beta}_{\mu\alpha}\equiv
0,$$  where $\Gamma^{\beta}_{\nu\alpha}$ is the Christofell symbol, which is
given by
$\Gamma^{\beta}_{\nu\alpha}=\frac{\chi}{2}{\big\lbrace}\hbox{\Large
a}^{\beta}(f'_{\nu},V'_{\alpha})+\frac{V^{\beta}}{2}(f'_{\nu}+V'_{\nu},\hbox{\Large a}_{\alpha})-
f^{\beta}(V'_{\nu},\hbox{\Large a}_{\alpha}){\big\rbrace}$, as
$g^{\mu\nu}f'_{\nu}=f^{\mu}-
\frac{\chi}{2}V^{\mu}\;,$ $\;g^{\mu\nu}V'_{\nu}=V^{\mu}(1-\frac{\chi}{2})\;,$
$\;\chi^{2}=\chi$,
and
 $\hbox{\Large a}=\dot{V}:=\frac{dV}{d\tau}$.
\noindent Therefore, from
$f^{\mu}:=\frac{(x-z)^{\mu}}{\Delta\tau}=\frac{R^{\mu}}{\rho}$ and
$\Delta\tau=\rho:=-V.\eta.(x-z)=- \frac{1}{1+\chi}V.g.(x-z)$, we find that
\begin{eqnarray}
D_{\alpha}\rho &=&\nabla_{\alpha}\rho=\lbrace {\hbox{\Large a}}_{f}\;\rho\;
f'_{\alpha}+f'_{\alpha}-V'_{\alpha},\rbrace\\
D_{\alpha}A^{\mu}&=&-\frac{1}{\rho}\lbrace (\hbox{\Large a}^{\mu}+{\hbox{\Large
a}}_fV^{\mu})f'_{\alpha}\rbrace-\frac{(f'_{\alpha}-
V'_{\alpha})V^{\mu}}{\rho^{2}}+\frac{V^{\beta}}{\rho}\Gamma^{\mu}_{\alpha\beta}\\
D_{\nu}A_{\mu}=g_{\mu\beta}D_{\nu}A^{\beta}&=&-\frac{f'_{\nu}\hbox{\Large
a}_{\mu}}{\rho}-
\frac{1+\chi}{\rho}V'_{\mu}(f_{\nu}{\hbox{\Large
a}}_f\;\rho+f'_{\nu}-V'_{\nu})-
\frac{\chi}{2\rho}{\big\lbrace} [(\hbox{\Large a}_{\mu},f'_{\nu}]+(\hbox{\Large
a}_{\mu},V'_{\nu}){\big\rbrace};
\end{eqnarray}
\noindent as $V^{\beta}g_{\beta\mu}=(1+\chi)V'_{\mu}\;$,
$\;f^{\beta}g_{\beta\mu}=f'_{\mu}+\chi V'_{\mu}\;,$ $\;\hbox{\Large
a}^{\beta}g_{\beta\mu}=\hbox{\Large a}_{\mu}\;,$
$\;g_{\nu\beta}V^{\alpha}\Gamma^{\beta}_{\alpha\beta}=-\frac{\chi}{2}\lbrace[(\hbox{\Large
a}_{\mu},f'_{\nu}]+(\hbox{\Large a}_{\mu},V'_{\nu})\rbrace\;$ and where we are
using the
short notation $\;(A,B):=AB+BA,\;$  $\;[A,B]:=AB-BA\;$ and ${\hbox{\Large
a}}_f:=\hbox{\Large a}.f$\\
 The retarded Maxwell field, ${F_{\mu\nu}}_{ret}:=D_{[\nu}A_{\mu]}$, is given
by
\begin{equation}
\label{Fl}
{F_{\mu\nu}}_{ret}=\frac{1+\chi}{\rho^{2}}[f'_{\mu},V'_{\nu}+\rho(\hbox{\Large
a}_{\nu}+\hbox{\Large
a}_{f}V'_{\nu})]=(1+\chi){\Big\lbrace}\frac{1}{\rho}[f'_{\mu},\hbox{\Large
a}_{\nu}]+\frac{\hbox{\Large
a}_{f}}{\rho}[f'_{\mu},V'_{\nu}]+\frac{[f'_{\mu},V'_{\nu}]}{\rho^{2}}{\Big\rbrace}
\end{equation}
\begin{equation}
\label{F}
F^{\mu\nu}_{ret}=\frac{1}{\rho^{2}}{\Big\lbrace}[f^{\mu}-\frac{\chi}{2}
V^{\mu},(1+\chi)\hbox{\Large a}^{\nu}\rho+V^{\nu}(1+\rho{\hbox{\Large
a}_{f}}){\Big\rbrace},
\end{equation}
 It is more convenient to work with covariant indices. So, in this paper, where
not explicitly
shown, as in (\ref{t}-\ref{t4}), below, we will be working with covariant
indices.

\noindent Using (\ref{Fl}) in
$4\pi\Theta_{\mu\nu}=F_{\mu\beta}g^{\alpha\beta}F_{\alpha\nu}-
g_{\mu\nu}\frac{F^{\alpha\beta}F_{\alpha\beta}}{4},$  we find
\begin{equation}
\label{t}
4\pi\rho^{4}\Theta=(1+3\chi){\Big\lbrace}[f',\rho\hbox{\Large
a}+V'(1+\rho{\hbox{\Large
a}}_f)].g.[f',\rho\hbox{\Large a}+V'(1+\rho{\hbox{\Large
a}}_f)]-\frac{g}{4}[f',\rho\hbox{\Large
a}+V'(1+\rho{\hbox{\Large a}}_f)]^{2}{\Big\rbrace}
\end{equation}
or $\Theta=\Theta_{2}+\Theta_{3}+\Theta_{4}$, with
\begin{equation}
\label{t2}
4\pi\rho^{2}\Theta_{2}=(1+3\chi){\Big\lbrace}[f',\hbox{\Large
a}+V'{\hbox{\Large
a}}_f].g.[f',\hbox{\Large a}+V'{\hbox{\Large a}}_f]-\frac{g}{4}[f',\hbox{\Large
a}+V'{\hbox{\Large a}}_f]^{2}{\Big\rbrace}
\end{equation}
\begin{equation}
\label{t3}
4\pi\rho^{3}\Theta_{3}=(1+3\chi){\Big\lbrace}[f',V'].g.[f',\hbox{\Large
a}+V'{\hbox{\Large
a}}_f]+[f',\hbox{\Large a}+V'{\hbox{\Large
a}}_f].g.[f,V']-\frac{g}{2}Tr[f',V'].g.[f',\hbox{\Large
a}]{\Big\rbrace}
\end{equation}
\begin{equation}
\label{t4}
4\pi\rho^{4}\Theta_{4}=(1+3\chi){\Big\lbrace}[f',V'].g.[f',V']-\frac{g}{2}[f',V']^{2}{\Big\rbrace}
\end{equation}
 It is worth to explicitly write (\ref{t}-\ref{t4}) and make some comments.
\begin{equation}
\label{t2e}
4\pi\rho^{2}\Theta_{2\;\mu\nu}=(1+\chi){\Big\lbrace}-
f'_{\mu}f'_{\nu}{\big\lbrace}{(1+\chi)\hbox{\Large a}}^{2}-{{\hbox{\Large
a}}_f}^{2}{\big\rbrace}+(f'_{\mu},T_{\nu}){\hbox{\Large
a}}_f\chi-T_{\mu}T_{\nu}\{\chi+f^{2}
(1+\chi)\}-\frac{g_{\mu\nu}}{2}\Bigl(\chi({{\hbox{\Large
a}}_f}^{2}-{\hbox{\Large a}}^{2})-
f^{2}\bigl((1+\chi){\hbox{\Large a}}^{2}-{{\hbox{\Large
a}}_f}^{2}\bigr)\Bigr){\Big\rbrace},
\end{equation}
\begin{equation}
\label{t3e}
4\pi\rho^{3}\Theta_{3\;\mu\nu}=(1+\chi){\Big\lbrace}2f'_{\mu}f'_{\nu}{\hbox{\Large a}}_f -
(f'_{\mu},T_{\nu})+\chi{\hbox{\Large
a}}_f(V'_{\mu},f'_{\nu})-(V'_{\mu},T_{\nu})
({\big\lbrace}(\chi+(1+\chi)f^{2}{\big\rbrace} -g_{\mu\nu}f^{2}{{\hbox{\Large
a}}_f}^{2}{\Big\rbrace},
\end{equation}
\begin{equation}
\label{t4e}
4\pi\rho^{4}\Theta_{4\;\mu\nu}=(1+\chi){\Big\lbrace}f'_{\mu}f'_{\nu} -
(f'_{\mu},V'_{\nu})-V'_{\mu}V'_{\nu}
{\big\lbrace}\chi+(1+\chi)f^{2}{\big\rbrace}
-\frac{g_{\mu\nu}}{2}(1+f^{2}){\Big\rbrace},
\end{equation}
where $T_{\mu}$ is a short notation for ${\hbox{\Large
a}}_{\mu}+V'_{\mu}{\hbox{\Large
a}}_f;$ then $f.T\equiv0.$\\ For $\chi=0\;\;$ (or, equivalently,  $\rho>0\;$),
(\ref{t2e}-\ref{t4e}),
with $f^{2}=0,$ coincide with their well known expressions$^{(10)}$ in the
usual formalism. In
particular, we have
\begin{equation}
\label{C}
f_{\mu}\Theta_{2}^{\mu\nu}{\bigg|}_{\rho>0}=f_{\mu}\Theta_{3}^{\mu\nu}{\bigg|}_{\rho>0}=0
\end{equation}
 as it must be for a radiation term.  $\Theta_{2}$ is the radiated portion of
$\Theta.$\\
But the terms with $f^{2},$ even with $f^{2}=0,$ cannot be dropped from the
above
equations, since they are necessary for producing the correct limits when
 $\rho\Longrightarrow0,$ or $x\Longrightarrow z.$ As $f^{\mu}:=\frac{(x-
z)^{\mu}}{\Delta\tau}=\frac{R^{\mu}}{\rho}$, in this limit  we have a
$0/0$-type of
indeterminacy, which can be raised with the L'Hospital rule and
$\frac{\partial}{\partial\tau}$
(more specifically, $\frac{\partial}{\partial\tau_{z}},$ as
$\Delta\tau=\tau_{x}-\tau_{z},$ and so,
$\frac{\partial\Delta\tau}{\partial\tau_{z}}=-1,\;\;\; \frac{\partial
R}{\partial\tau}=-V,$ etc).
Therefore,
$$\lim_{R\to0}f=V,$$
and$$\lim_{R\to0}f'_{\mu}f^{\mu}=\lim_{R\to0}\frac{R.\eta.R}{\rho^{2}}=-
1.$$
To find the limit of something when $\rho\rightarrow0$ will be done so many
times in this
paper that it better be systematized.   We want to find
\begin{equation}
\label{LR}
\lim_{R\to0}\frac{N(R)}{\rho^{n}},
\end{equation}
where $N(R)$ is a homogeneous function of R,  $N(R){\Big|}_{R=0}=0$. Then, we
have to
apply the L'Hospital rule consecutively until the indeterminacy is resolved.
As
$\frac{\partial\rho}{\partial\tau}=-(1+\hbox{\Large a}.R)$, the denominator of
(\ref{LR}) at
$R=0$ will be different of zero only after the $n^{th}$-application of the
L'Hospital rule, and
then, its value will be $(-1)^{n}n!$\\
If p is the smallest integer such that $N(R)_{p}{\Big|}_{R=0}\neq0,$ where
$N(R)_{p}:=\frac{d^{p}}{d{\tau}^{p}}N(R)$, then
\begin{equation}
\label{NR}
\lim_{R\to0}\frac{N(R)}{\rho^{n}}=\cases{\infty,& if $p<n$\cr
              (-1)^{n}{\frac{N(0)_{p}}{n!}},& if $p=n$\cr
0,& if $p>n$\cr}
\end{equation}
\begin{itemize}
\item Example 1: $\cases{f=\frac{R}{\rho}.& $n=p=1 \Longrightarrow
\lim_{R\to0}f=V$\cr
f^{2}=\frac{R.\eta.R}{\rho^{2}}.&$ n=p=2 \Longrightarrow
\lim_{R\to0}f^{2}=-1.$\cr}$
\item Example 2: $\frac{[f'_{\mu},\hbox{\Large
a}_{\nu}]}{\rho}=\frac{[R_{\mu},\hbox{\Large
a}_{\nu}]}{\rho^{2}}\;\;$ $\;\Longrightarrow \;$ $\;p=1<n=2\Longrightarrow
\lim_{R\to0}\frac{[f'_{\mu},\hbox{\Large a}_{\nu}]}{\rho}$ diverge
\item Example 3: $\frac{{\hbox{\Large
a}}_f}{\rho}f^{[\mu}V^{\nu]}=-\frac{\hbox{\Large
a}.R}{\rho^{3}}R^{[\mu}V^{\nu]}\Longrightarrow p=4>n=3$ \qquad
$\lim_{\rho\to0}\frac{{\hbox{\Large a}}_f}{\rho}f^{[\mu}V^{\nu]}=0$
\item Example 4
$\frac{[f_{\mu},V'_{\nu}]}{\rho^{2}}=\frac{[R_{\mu},V'_{\nu}]}{\rho^{2}}\;\;
\;\Longrightarrow
\;\;p=2<n=3\Longrightarrow\lim_{R\to0}\frac{[f_{\mu},V'_{\nu}]}{\rho^{2}}$
diverge
\end{itemize}
The second term of the RHS of (\ref{Fl}) does not contribute to the electron
self-field at
$\rho=0,$  but the first and the third terms  diverge, as expected, although
they produce
integrable contributions to the electron self field stress tensor, as we show
now. Let us find
the integral of the electron self field stress tensor at the electron
causality-line:
$\lim_{\rho\to0}\int dx^{4}\Theta,$  or $\lim_{\rho\to0}\int d\tau\rho^{2}d\rho
d^{2}\Omega\Theta,$ in terms of retarded coordinates$^{9,13,14}$,
$x^{\mu}=z^{\mu}+\rho
f^{\mu},$ where $d^{2}\Omega$ is the element of solid angle in the charge
rest-frame.\\ But,
we will first prove a useful result from (\ref{NR}), when the numerator has the
form
$N_{0}=A_{0}.g_{0}.B_{0},$ where A and B represent two possibly distinct
homogeneous
functions of R, and the subindexes indicate the order of $\frac{d}{d\tau}$.
Then
$$N_{1}=(A_{1}.g.B_{0}+A_{0}.g.B_{1})+A_{0}.g_{1}.B_{0};$$
$$N_{2}=(A_{2}.g.B_{0}+2A_{1}.g.B_{1}+A_{0}.g.B_{2})+(A_{1}.g_{1}.B_{0}+A_{0}.g_{1}.B_{1
})+A_{0}.g_{2}.B_{0};$$
and, generically,
\begin{equation}
\label{NP}
N_{p}=\sum_{a=0}^{p}\pmatrix{p \cr
a\cr}A_{p-a}.g.B_{a}+\sum_{a=0}^{p-1}\pmatrix{p\cr
a\cr}A_{p-a-1}.g_{1}.B_{a}+\dots=\sum_{b=0}^{p}\sum_{a=0}^{p-b}\pmatrix{p-b\cr
a\cr}A_{p-
a-b}.g_{b}.B_{a}
\end{equation}
As, for the cases we are considering in this paper, the metric g is independent
of R, b in
(\ref{NR}) will always be zero.
So, for applying (\ref{NR}) in this case, we just have to find the
$\tau$-derivatives  of A and B
that produce the first non- null term at the limit of $R\rightarrow0$.\\
Applying (\ref{NR}) and (\ref{NP}) for finding $\lim_{\rho\to0}\int
dx^{4}\Theta$ we just have
to consider the first term of the RHS of (\ref{t}); the second one, as the
trace of the first, will
have a similar behaviour.
\begin{itemize}
\item Example 5
$$\lim_{\rho\to0}\frac{\rho^{2}[f',\rho\hbox{\Large a}+V'(1+\rho{\hbox{\Large
a}}_f)].g.[f',\rho\hbox{\Large a}+V'(1+\rho\;{\hbox{\Large
a}}_f)]}{\rho^{4}}=$$
$$=\lim_{\rho\to0}\frac{[R,\rho\hbox{\Large a}+V'(1+\hbox{\Large
a}.R)].g.[R,\rho\hbox{\Large
a}+V'(1+\hbox{\Large a}.R)]}{\rho^{4}}$$

As $$A_{0}=B_{0}=[R,\rho\;\hbox{\Large a}+V'(1+\rho\hbox{\Large
a}.R)]\Longrightarrow
A_{2}=B_{2}=[\hbox{\Large a},V']+{\cal O}(R)$$
Therefore, according to (\ref{NP}), for producing a non null $N_{p}$, a and p
(as $b=0$) must
be given by
$$p-a= a=2\Longrightarrow p=4=n\Longrightarrow N_{4}=6[\hbox{\Large
a},V'].g.[\hbox{\Large a},V']+{\cal O}(R).$$
So, from (\ref{NR}),

$$\lim_{\rho\to0}\frac{\rho^{2}[f',\rho\hbox{\Large a}+V'(1+\rho{\hbox{\Large
a}}_f)].g.[f',\rho\hbox{\Large a}+V'(1+\rho{\hbox{\Large
a}}_f)]}{\rho^{4}}=\frac{1}{4}[\hbox{\Large a},V'].g.[\hbox{\Large a},V']$$
Therefore, from (\ref{t}), with $\chi=1$,
\begin{equation}
\label{te}
\lim_{\rho\to0}\int dx^{4}\;\Theta=[\hbox{\Large a},V'].g.[\hbox{\Large a},V']-
\frac{g}{4}[\hbox{\Large a},V']^{2}
\end{equation}
It is interesting that (\ref{te}) comes entirely from the velocity term,
$\frac{[f',V']}{\rho^{2}}$,
as we can see from the following example.
\item Example 6
$$\lim_{\rho\to0}\frac{\rho^{2}[f',V'].g.[f',V']}{\rho^{4}}=\lim_{\rho\to0}\frac{[R,V].g.[R,V']}{\rho^{
4}}=\frac{1}{4}[\hbox{\Large a},V'].g.[\hbox{\Large a},V']$$
as $A_{2}=B_{2}=[\hbox{\Large a},V']+[R',\dot {\hbox{\Large a}}]\Longrightarrow
p-
a=a=2\Longrightarrow N_{4}=6[\hbox{\Large a},V']+{\cal O}(R)\;\;$ and $ p=n=4.$
So,
$$\lim_{\rho\to0}\int dx^{4}\Theta=\lim_{\rho\to0}\int dx^{4}\Theta_{4}$$
The other contributions just cancel to zero,
$$\lim_{\rho\to0}\int dx^{4}\Theta_{2}=-\lim_{\rho\to0}\int dx^{4}\Theta_{3}.$$
\end{itemize}

\section{THE NEW EQUATION OF MOTION}
Let us now derive the electron equation of motion, which can be obtained from
$$\lim_{\varepsilon\to0}\int
dx^{4}D_{\nu}T^{\mu\nu}\theta(\rho-\varepsilon)=0,$$
where $T^{\mu\nu}$ is the total electron energy-momentum tensor, which includes
the
contribution from the electron kinetic energy, from its interaction with
external fields and from
its self field. Let us move directly to the part that will produce novel
results:
\begin{equation}
\label{eqm}
m\int a^{\mu}d\tau=\int F_{ext}^{\mu}d\tau-\lim_{\varepsilon\to0}\int
dx^{4}D_{\nu}\Theta^{\mu\nu}\theta(\rho-\varepsilon),
\end{equation}
\noindent where $F_{ext}^{\mu}$ is the external forces acting on the electron,
and the last
term represents the impulse carried out by the emitted electromagnetic field in
the Bhabha
tube surrounding the electron worldline, defined by the Heaviside function,
$\theta(\rho-
\varepsilon)$.  Using
$D_{\nu}\Theta^{\mu\nu}=\nabla_{\nu}\Theta^{\mu\nu}+\Theta^{\alpha\nu}\Gamma^{\mu}_{\alpha\nu},$ as $\Gamma^{\nu}_{\alpha\nu}=0,$ and the divergence theorem, we have that the
last term of the RHS of (\ref{eqm}) is transformed into
\begin{equation}
\label{gauss}
\lim_{\varepsilon\to0}\int
dx^{4}\Biggl\{\Theta^{\alpha\nu}\Gamma^{\mu}_{\alpha\nu}
\theta(\rho-\varepsilon)-\Theta^{\mu\nu}\nabla_{\nu}\rho\;\delta(\rho-\varepsilon)\Biggr\}.
\end{equation}
\noindent The last term of this equation represents the flux of 4-momentum
through the
cylindrical hypersurface $\rho=\varepsilon.$ Let us denote it by $P^{\mu}$:
\begin{equation}
\label{P}
P^{\mu}=\lim_{\varepsilon\to0}\int
dx^{4}\Theta^{\mu\nu}\nabla_{\nu}\rho\;\delta(\rho-
\varepsilon),
\end{equation}
As $\nabla_{\nu}\rho= \rho {\hbox{\Large a}}_ff'_{\nu}+f'_{\nu}-V'_{\nu},$ and
$\Theta=\Theta_{2}+\Theta_{3}+\Theta_{4},$ we can write
$P^{\mu}:=P^{\mu}_{0}+P^{\mu}_{1}+P^{\mu}_{2},$ with
\begin{equation}
\label{P2}
P^{\mu}_{2}=\lim_{\varepsilon\to0}\int
dx^{4}\Theta^{\mu\nu}_{4}(f'-V')_{\nu}\;\delta(\rho-
\varepsilon),
\end{equation}
\begin{equation}
\label{P1}
P^{\mu}_{1}=\lim_{\varepsilon\to0}\int dx^{4}\lbrace
\Theta^{\mu\nu}_{4}f'_{\nu}\;\rho\;{\hbox{\Large a}}_f+\Theta^{\mu\nu}_{3}(f'-
V')_{\nu}\rbrace\;\delta(\varepsilon-\rho),
\end{equation}
\begin{equation}
\label{P0}
P^{\mu}_{0}=\lim_{\varepsilon\to0}\int dx^{4}\lbrace
\Theta^{\mu\nu}_{3}f'_{\nu}\rho\hbox{\Large a}_{f}+\Theta^{\mu\nu}_{2}(f'-
V')_{\nu}\rbrace\;\delta(\rho-\varepsilon),
\end{equation}
\noindent $P^{\mu}_{1}$ and $P^{\mu}_{2}$ are both null. In order to show this
we need to
apply (\ref{NR}) for a N(R) that has a generic form, $N=A.g.B\;C$, where A,B,
and C are
functions of R, such that $N(R=0)=0$. Then it is easy to show, from (\ref{NP}),
that
\begin{equation}
\label{AgBC}
N_{p}=\sum_{b=0}^{p}\sum_{a=0}^{p-b}\sum_{c=0}^{a}\pmatrix{p-b\cr
a\cr}\pmatrix{a\cr
c\cr}A_{p-a-b}.\;g_{b}\;.B_{a-c}\;C_{c}
\end{equation}
\begin{itemize}
\item From (\ref{t4}), the integrand of (\ref{P2}), produces (again, we do not
need to consider
the trace term)
$$\lim_{\rho\to0}\frac{\rho^{2}[f',V'].g.[f',V'].(f-
V)}{\rho^{4}}=\lim_{\rho\to0}\frac{[R,V'].g.[R,V'].g.(R-\rho V)}{\rho^{5}},$$
or, schematically
$$\lim_{\rho\to0}\frac{A.g.A\;C}{\rho^{5}}$$
with $A_{0}=B_{0}=[R,V],$ and $C_{0}=(R-V\rho).$ Then, $A_{2}=[a,V]+{\cal
O(R)},\;\;$
$\;C_{2}=a+{\cal O(R)},$ and we have, from (\ref{AgBC}),  the following
restrictions on a,b,c
for producing a $N(R=0)_{p}\neq0:$

$c=2;$\quad$b=0$ (always, as g does not depend on R); $a-c=2;$ and $p-a=2$ or
$p=6>n=5.$
Therefore, according to (\ref{NR})
\begin{equation}
\label{P20}
P^{\mu}_{2}=0
\end{equation}
\item From the second term of the integrand of (\ref{P1}) and from (\ref{t3})
we have
$$\lim_{\rho\to0}\frac{\rho^{2}[f',V'].g.[f',\hbox{\Large a}+V'{\hbox{\Large
a}}_f].(f-
V)}{\rho^{3}}=\lim_{\rho\to0}\frac{[R,V'].g.[R,\rho\;\hbox{\Large
a}+V'\hbox{\Large a}.R].(R-
\rho\;V)}{\rho^{5}}.$$
So, from (\ref{AgBC}) with
 $$C_{0}=R-V\rho\;\;\Longrightarrow C_{2}={\hbox{\Large a}}+{\cal
O(R)}\Longrightarrow
c=2.$$
$$B_{0}=[R,V]\;\;\Longrightarrow B_{2}=[\hbox{\Large a},V]+{\cal
O(R)}\Longrightarrow
a=4.$$
$$A_{0}=[R,{\hbox{\Large a}}]\;\;\Longrightarrow A_{1}=[{\hbox{\Large
a}},V]+{\cal
O(R)}\Longrightarrow p=5=n.$$
But $[\hbox{\Large a},V].g.[\hbox{\Large a},V].{\hbox{\Large a}}=(1-
\frac{\chi}{2}){\hbox{\Large a}}^{2}{\hbox{\Large a}}$ and it is cancelled by
the contribution
from the trace term of (\ref{t3}).
Therefore, $$\lim_{\rho\to0}\rho^{2}\Theta^{\mu\nu}_{3}(f'-V')_{\nu}=0$$
\item From the first term of the integrand of (\ref{P1}) and from (\ref{t4}) we
have
$$\lim_{\rho\to0}\frac{\rho^{2}[f',V'].g.[f',V']gf'\rho{\hbox{\Large a}}.f
}{\rho^{4}}=\lim_{\rho\to0}\frac{[R,V'].g.[R,V']gR{\hbox{\Large a}}.R
}{\rho^{5}}.$$
So, from (\ref{AgBC}), with $$C=C_{0}=ggR\;\hbox{\Large a}.R\;\;\Longrightarrow
C_{3}=-
3V{\hbox{\Large a}}^{2}+{{\cal O(R)}}\Longrightarrow c=3.$$
$$A=B=B_{0}=[R,V]\;\;\Longrightarrow A_{2}= B_{2}=[\hbox{\Large a},V]+{{\cal
O(R)}}\Longrightarrow a=5 \hbox{ and }\;p=7>n=5.$$
Therefore, from (\ref{NR})
$$\lim_{\rho\to0}\rho^{2}\Theta^{\mu\nu}_{4}f'_{\nu}\rho\;{\hbox{\Large
a}}_f=0$$
\end{itemize}
Consequently
 \begin{equation}
\label{P20}
P^{\mu}_{1}=0
\end{equation}
and
\begin{equation}
\label{P0F}
P^{\mu}=P^{\mu}_{0}
\end{equation}
\noindent $P^{\mu}_{0}$ is distinguished from  $P^{\mu}_{1}$ and $P^{\mu}_{2}$
for not
being $\rho$-dependent. This has 3 important implications:
\begin{itemize}
\item It is not necessary to use the L'Hospital rule on its determination;
\item It is, therefore, not affected by the limit of $\varepsilon\rightarrow0$;
\item The presence of $\delta(\rho-\varepsilon)\;$ with $\varepsilon>0$
requires $\chi=0$,
\end{itemize}
which is in agreement with (\ref{C}).
The physical meaning of this is that the flux of 4-moment through the
cylindrical surface
$\rho=\varepsilon$ comes entirely from the photon field and it requires,
therefore, $\chi=0$.
But we will keep $\chi$, as  $\chi'$ through the calculation just to see what
would be its
contribution to $P^{\mu}_{0}$. We just have to make $\chi'=0,$ at the end.
 It is convenient now to introduce a spacelike 4-vector N, defined by
$N^{\mu}:=(f-V)^{\mu}$
and such that $N.\eta.V=N.g.V=0$ and $N.g.N=N.\eta.N=1$. N satisfies
\begin{equation}
\label{n1}
\frac{1}{4\pi}\int d\Omega \overbrace{N\cdots N}^{odd\;number}=0
\end{equation}
\begin{equation}
\label{n2}
\frac{1}{4\pi}\int d\Omega NN=\frac{\eta+VV}{3};
\end{equation}
\noindent as can be found, for example, in references [4,13] or in the
appendix B of of reference [10]. From (\ref{t3}) we have
\begin{equation}
\label{t3f}
\int  d^{4}x\Theta^{\mu\nu}_{3}f'_{\nu}\rho{\hbox{\Large
a}}_{f}=\frac{\chi'}{4\pi}\int d\tau
d^{2}\Omega({\hbox{\Large a}}.N)^{2}N^{\mu}=0,
\end{equation}
{}From (\ref{t2e}), we have
\begin{equation}
4\pi\rho^{2}\Theta^{\mu\nu}_{2}N_{\nu}=(1+\chi'){\biggl (}({\hbox{\Large
a}.N})^{2}-
{\hbox{\Large a}}^{2}{\biggr )}{\biggl (}V^{\mu}+(1+\chi')N^{\mu}{\biggr )},
\end{equation}
which, with (\ref{n1}) and (\ref{n2}), gives
\begin{equation}
\label{tN}
\lim_{\rho\to0}\int  d^{4}x\Theta_{2}^{\mu\nu}N_{\nu}=-\int
d\tau\frac{2(1+\chi')}{3}a^{2}V^{\mu}.
\end{equation}
Then, from (\ref{tN}), (\ref{t3f}) and (\ref{P0F}), we have
\begin{equation}
\label{Larmor}
P^{\mu}=-\int d\tau\frac{2(1+\chi')}{3}a^{2}V^{\mu}.,
\end{equation}
As we have to make $\chi'=0,$ it is just the Larmor term. It is interesting
that the contribution
from a $\chi\neq0$ would be just to make it to be     twice its usual value. \\

In order to get to the electron equation of motion we still need to work out
the first part of
(\ref{gauss}). In contradistinction to the second part, this one comes
multiplied by
$\Theta(\rho-\varepsilon)$ and not by $\delta(\rho-\varepsilon.)$ Therefore, it
has no part that
is independent of $\rho$ (and of $\chi$). Its limit when $\rho\rightarrow0$ can
be evaluated
with (\ref{AgBC}) and with
$g^{\rho\alpha}g^{\sigma\beta}\Gamma^{\mu}_{\alpha\beta}=\frac{\chi}{4}{\biggl(}{\hbox{\Large a}}^{\mu}(V^{\rho},f^{\sigma})-f^{\mu}({\hbox{\Large
a}}^{\rho},V^{\sigma})+V^{\mu}({\hbox{\Large
a}}^{\rho},f^{\sigma})-{\hbox{\Large
a}}^{\mu}V^{\rho}V^{\sigma}{\biggr)}:$

$$\lim_{\rho\to0}\frac{\rho^{2}[f',\rho\hbox{\Large a}+V'(1+\rho{\hbox{\Large
a}}_f)].g.[f',\rho\hbox{\Large a}+V'(1+\rho\;{\hbox{\Large
a}}_f)].{\biggl(}\hbox{\Large
a}(f,V)+V(\hbox{\Large a},f)-f(\hbox{\Large a},V)-\hbox{\Large a}VV{\biggr
)}\frac{\chi}{4}}{\rho^{4}}=$$
$$=\lim_{\rho\to0}\frac{[R,\rho\hbox{\Large a}+V'(1+\hbox{\Large
a}.R)].g.[R,\rho\hbox{\Large
a}+V'(1+\hbox{\Large a}.R)].{\biggl (}\hbox{\Large a}(R,V)+V(\hbox{\Large
a},R)-
R(\hbox{\Large a},V)-\rho\hbox{\Large a}VV {\biggr
)}\frac{\chi}{4}}{\rho^{5}},$$
with $$C_{0}=\hbox{\Large a}(R,V)+V(\hbox{\Large a},R)-R(\hbox{\Large
a},V)-\rho\;
\hbox{\Large a}VV\Longrightarrow C_{1}=-\hbox{\Large a}VV+{\cal
O}(R)\Longrightarrow
c=1,$$
and, like in the example 5, $$A_{2}=B_{2}=[\hbox{\Large a},V']+{\cal
O}(R)\Longrightarrow a-
c=2 \hbox{ and } p-a=2\Longrightarrow a=3\;{\hbox{and}}\;p=5=n$$
Therefore, $$N(R=0)_{5}=\frac{1}{5!}\pmatrix{5\cr 3\cr}\pmatrix{3\cr 1\cr}[
\hbox{\Large a},V'].g.[
\hbox{\Large a},V'].(VV)\hbox{\Large a}\frac{\chi}{4}.$$
Then,
\begin{equation}
\label{TG}
\lim_{\varepsilon\to0}\int dx^{4}\Theta^{\alpha\nu}\Gamma^{\mu}_{\alpha\nu}
\theta(\rho-
\varepsilon)=\frac{(1+3\chi)}{2}\int d\tau{\Biggl(}[\hbox{\Large a}_{\alpha},
V_{\rho}]g^{\rho\sigma}[\hbox{\Large a}_{\sigma},V_{\beta}]-
g_{\alpha\beta}\frac{[\hbox{\Large a},V]^{2}}{4}{\Biggr)}\frac{\chi}{4}
V^{\alpha}V^{\beta}\hbox{\Large a}^{\mu}=-\frac{\chi}{4}\int d\tau{\hbox{\Large
a}}^{2}\hbox{\Large a}^{\mu}.
\end{equation}
This term corresponds to the energy associated to the electron manifold
curvature. Observe
that it is proportional to $\chi.$\\
Finally, from (\ref{eqm}), (\ref{gauss}), (\ref{P}), (\ref{Larmor}) and
(\ref{TG}),  we can write
the electron equation of motion, obtained from the Lienard-Wiechert solution,
as
\begin{equation}
\label{finally}
(m-\frac{1}{4}\;{\hbox{\Large a}}^{2})\hbox{\Large
a}^{\mu}=F^{\mu}_{ext}-\frac{2{\hbox{\Large
a}}^{2}}{3}V^{\mu},
\end{equation}

 The external force provides the work for changing the charge velocity, for the
energy
dissipated by the radiation, and for the curvature of the electron manifold.
The correctness of
the presence of the ${\hbox{\Large a}}^{2}\hbox{\Large a}^{\mu}$ term in the
above equation
must be decided through experimental means. It can, in principle, be detected
in synchrotron
accelerators, despite the difficulties caused by the smallness of
${\hbox{\Large a}}^{2}$ face
the experimental uncertainties$^{(15)}$.
 Indirect evidences would be the analysis of the theoretical implications of
this term in
(\ref{finally}).  This is left for future works.\\

\begin{center}
Energy conservation
\end{center}
Eq. (\ref{finally}) is non-linear, like the Lorentz-Dirac equation, but it does
not contain the
Schott term, the responsible for its spurious behaviour. This is good since it
signals that there
will be no problem with causality violation.  But the Schott term in the LDE
has also the role
of giving the guaranty of energy conservation, which is obviously missing in
(\ref{finally}).
Assuming that the external force is of electromagnetic origin,
($F_{ext}^{\mu}=F_{ext}^{\mu\nu}V_{\nu}),$ the contraction of V with eq.
(\ref{finally}) would
require a contradictory $a^{2}\equiv0$. But this is just an evidence that
(\ref{finally}) cannot
be regarded as a fundamental equation. It would be better represented as
\begin{equation}
\label{ffinally}
m\hbox{\Large a}^{\mu}-\frac{1}{4}\;<{\hbox{\Large a}}^{2}\hbox{\Large
a}^{\mu}>=F^{\mu}_{ext}-<\frac{2}{3}{\hbox{\Large a}}^{2}V^{\mu}>,
\end{equation}
with $$<\frac{2}{3}{\hbox{\Large
a}}^{2}V^{\mu}>=P^{\mu}=\lim_{\varepsilon\to0}\int
dx^{4}\Theta^{\mu\nu}\nabla_{\nu}\rho\;\delta(\rho-\varepsilon),$$
$$<\frac{1}{4}{\hbox{\Large a}}^{2}{\hbox{\Large
a}}^{\mu}>=\lim_{\varepsilon\to0}\int
dx^{4}\Theta^{\alpha\nu}\Gamma^{\mu}_{\alpha\nu}    \theta(\rho-\varepsilon).$$
It is just an effective or average result, in the sense that the contributions
from the electron
self field must be calculated, as in (\ref{eqm}), by the electromagnetic
energy-momentum
content of a spacetime volume containing the charge causality-line, in the
limit of
$\rho\rightarrow0$:
\begin{equation}
\label{ec}
m\int {\hbox{\Large a}}.Vd\tau=\int F_{ext}.Vd\tau-\lim_{\varepsilon\to0}\int
dx^{4}f_{\mu}D_{\nu}\Theta^{\mu\nu}\theta(\rho-\varepsilon).
\end{equation}
Observe that in the last term,V, the speed of the electron, is replaced by f
the speed of the
electromagnetic interaction; only in the limit of $R\rightarrow0$ is that
$f\rightarrow V.$
We have to repeat the same steps from (\ref{eqm}) to (\ref{finally}) in order
to calculate this
last term and to prove (done in the appendix) that it is null:
 \begin{equation}
\label{fec}
\lim_{\varepsilon\to0}\int
dx^{4}f_{\mu}D_{\nu}\Theta^{\mu\nu}\theta(\rho-\varepsilon)=0,
\end{equation}
So, there is no contradiction anymore. Besides, it throws some light on the
meaning of
$F_{\mu\nu}$, as we discuss in the following section.
\section{THE MEANING OF THE MAXWELL'S FIELDS}
With this new spacetime structures we have eliminated the non-integrable
singularities and
understood the problems with conservation of energy and causality violation
that plague the
old LDE, but there is still a remaining problem. How to conciliate this with
the unboundness of
$\Theta^{\mu\nu}$ as $\rho\rightarrow0$? This remaining singularity comes from
(\ref{1}), the
solution of the wave equation
\begin{equation}
\label{Dalembert}
\Box A^{\mu}=J^{\mu},
\end{equation}
 of the Minkowskian formalism. We must realize now, that we are still giving a
dichotomic
treatment to the electron and to its field. The electron is treated as a
classical particle, that is,
as a localized object in a well defined trajectory. When we use directional
derivatives
(\ref{directional}) and the induced metric (\ref{metric}), we are requiring
that the electron
follows a causality-cone generator of its (instantaneous) causality-cone, which
corresponds to
a local (point to point) causality implementation. But the LWS (\ref{1})
describes a wave front
propagating in the entire light-cone (not just one of its generator, as for the
electron) and this
corresponds to a global causality implementation.\\
By local implementation of causality we refer to the restriction on physical
objects to remain
in a causality-cone-generator, so that we know its location,  point after
point; while a global
implementation of causality just requires that a physical object be restricted
to a causality
cone, or in Minkowski term, on a light-cone if it is massless, or inside a
light-cone, if it is
massive.
One could argue that this global characteristic is something intrinsic to the
very nature of a
field. But it does not have to be this way. The concept of field, as a
distributed object, can be
compatible with the concept of photon, a localized object, if the field
equations have soliton-like solutions.

As far as the actually fundamental electromagnetic interaction is the one
represented by the
exchange of a single photon, it cannot be represented by the
Maxwell
electromagnetic field. It represents rather a kind of effective or average
interaction.
Consider, for example, the definitions contained in the Coulomb's law and in
the Gauss's law.
While the Coulomb's law defines the  {\it observed} electric force between two
point charges
as a vector physical manifestation acting on each charge, and that exists only
in the
direction given
by the straight line that passes by them, the Gauss's law describes the {\it
inferred} electric
field as existing around a single charge, independent of the presence of the
other charge.
The electric field, as it is well known, is extracted from the Gauss's law
through the
integration of its flux across a surface, having the appropriate symmetry, {\it
enclosing} the
charge,
\begin{equation}
\label{Gauss}
{\vec E}(x)={\hat e} \frac{\int^{x}_{V}\rho dv}{\int_{\partial V}dS}.
\end{equation}

The eq. (\ref{Gauss}) puts in evidence the effective or average character of
the Maxwell's
concept of field; it gives also a hint on the meaning and origin of the field
singularity. If the
electric field can be visualized in terms of exchanged photons, then according
to
(\ref{Gauss}), the frequency or the number of these exchanged photons must be
proportional
to the enclosed net charge. And if we take ${\vec E}$, as suggested by the
Gauss' law, as a
measure of the average number of photons emitted/absorbed by a point charge, we
can
schematically write, $E\sim \frac{n}{4\pi r^{2}},$ where n is the number of
photon per unit of
time crossing an spherical surface of radius r and centred on the charge. Then,
the
unbounded dependence of E with r when $r\rightarrow0$ does not represent a
physical fact
like an increasing number of photons, but just an increasing average number of
photons per
unit area, as the number of photons remains constant but the area tends to
zero. So, a field
singularity would have no physical meaning, it would just be a consequence of
this average
nature of the Maxwell's field.\\
In order to give a consistent classical treatment to the electron and to its
field in this
geometry, we have to consider the classical electromagnetic field as described
by a
punctiform classical photon propagating along a light-cone generator. Let us
represent this
classical photon by $A^{\mu}(x)_{f}$, where f labels its light-cone generator,
its causality-line.
This label f implies that the photon is causally constrained to remain in this
f-generator, that
is
\begin{equation}
\label{daf}
\partial_{\mu}A_{f}{\Big|}_{d\tau+f.dx=0}\equiv\nabla^{f}_{\mu}A_{f}.
\end{equation}
Its wave equation is then
\begin{equation}
\label{fMaxwell}
\Box_{f}A^{f}_{\mu}:=\nabla^{f}_{\alpha}\eta^{\alpha\beta}\nabla^{f}_{\beta}A^{f}_{\mu}=4\pi
J_{\mu},
\end{equation}
 which corresponds to the replacement, in the standard equation, of the usual
derivative,
$\partial_{\mu}$, by the directional derivative along the f-causality-cone
generator,
$\nabla^{f}_{\mu}$.\\
The differences between (\ref{Dalembert}) and (\ref{fMaxwell}) are very
significative. The
field A(x) represents a distributed wave propagating along every direction; the
operators
$\partial_{x},\;\partial_{y},\;\partial_{z},\;$and$\;\partial_{t}$ has, each
one, an equal
importance and are equally weighted in $\Box.\;\;$ $A^{\mu}(x)_{f}$ represents
a
localized
object propagating along a generator f; only the derivative along f is
important and is
considered in $\Box_{f}$. The use of A(x) constitutes an average or statistical
approach,
considering that the electromagnetic wave consist of a huge number of photons,
each one
propagating along a null direction f, and described by an $A^{\mu}(x)_{f}$.
 We can expect then, that an integration of $A^{\mu}(x)_{f}$ over all the
generators of its
light-cone reproduces the Minkowskian $A^{\mu}(x)$,
\begin{equation}
\label{classicfield}
 A(x)=\frac{1}{4\pi} \int d\Omega_{f}\; A(x,\tau,f),
\end{equation}
and that, the same integration over (\ref{fMaxwell}) reproduces
(\ref{Dalembert}). This is
possible because $A_{f}$ has an even$^{(2)}$ functional dependence on  f, and
so, the f-odd
term in (\ref{fMaxwell}) is integrated out to zero, by symmetry $(\int
d\Omega_{f}f\equiv 0)$.\\
The equation (\ref{fMaxwell}) is solved, in reference$^{(2)}$, for a fixed and
constant f,  by
the following Green function,
\begin{equation}
\label{expqgreen}
G(x,\tau)_{f}=-\frac{1}{2}\theta (-af.\Delta x)\theta(-a{\bar{f}}.\Delta x)
\delta (\Delta\tau+{f.\Delta x}),
\end{equation}
where $a=\pm1$.\\ $G(x,\tau)_{f}$ and its formalism have some remarkable
properties to be
fully discussed elsewhere$^{(2)}$ and from which we underline:
\begin{enumerate}
\item It is singularity-free. In contradistinction to (\ref{cG}), below, the
argument of the delta
function in (\ref{expqgreen}) is linear in $\Delta x$, which shows that the
``classical photon"
field propagates without changing its amplitude and its shape, like a soliton
or a classical
particle;
\item The Dirac's delta, $\delta (\Delta\tau+{f.\Delta x})$, together with the
overall global
causality-constraint (\ref{lcone}) implies that the photon, after being emitted
by a charge at
their common causality-cone vertex, freely propagates along a f-generator of
this  causality-
cone. Even though the physical description remains in $(3+1)$ dimensions, the
field
dynamics, described through (\ref{expqgreen}), is essentially $(1+1)$
dimensional, along the
space direction of propagation of the exchanged photon;
\item It is conformally invariant, which is the appropriate symmetry (not
shared by the usual
Green function, (\ref{cG}) below) for describing the photon propagation. It is
free of scale
dependent parameters $(f^{2}=0)$;
\item The constraint (\ref{hypercone}) replaces and generalizes (\ref{lcone}).
Both sides of
$(\Delta \tau)^{2}=(\Delta t)^{2}+(\Delta{\vec{x})}^{2}$ are invariant under
transformations of
the $SO(3,1),$ the invariance group of a Minkowski spacetime $({\vec{x}},t)$,
but in $(\Delta
t)^{2}=(\Delta\tau)^{2}+(\Delta{\vec{x})}^{2}$,
  both sides are invariant$^{(1)}$ under transformations of $O_{4},$ the
rotation group of a 4-
dimension Euclidean spacetime $({\vec{x}},\tau).$ The change of $({\vec{x}},t)$
to
$({\vec{x}},\tau)$ is a Wick rotation$^{(16,17)}$ without the need of an
uncomprehensible
imaginary time. Physically it means that, for an  Euclidean   4-dimensional
spacetime, events
should be labelled not with the time measured in the observer clock, but with
their
propertime, measured on their local clocks. $O_{4}$ is the  causality-cone
invariance under
rotation around its t-axis. Care must be taken with the interpretation of
transformations that
corresponds to rotation of the $\tau$-axis, as they involve Lorentz and
conformal
transformations;
\item Following the recipe (\ref{classicfield}), the integration of
$G(x,\tau)_{f}$ over the f-
directions, for $\Delta\tau=0$ and for a fixed x, reproduces$^{(2)}$ the
standard $G(x)$
\begin{equation}
\label{cG}
 G(x)=\frac{1}{4\pi} \int d\Omega_{f}\; G(x,\tau,f)=\frac{1}{2}\frac{\theta
(at)}{r}\;\delta(r+at)=\theta (at)\;\delta(r^{2}-t^{2}),
\end{equation}
This shows that the singularity at $r=0$ is not physical; it is merely a
consequence of the
average meaning of G(x) and of A(x), expressed in (\ref{classicfield}).
\item The f of $G(x-y)_{f}$, represents a generator of the causality-cone with
vertex at x, and
describes two types of propagating signals (see the figure):
\begin{enumerate}
\item $\Delta x^{0}=x^{0}-y_{1}^{0}>0.\;\;$ $G(x-y)_{f}$ describes the
propagation of a photon
emitted at
$y_{1},$ along f, and being observed at x. $J(y_{1})$ is its source.
\item $\Delta x^{0}=x^{0}-y_{2}^{0}<0.\;\;$ $G(x-y)_{\bar{f}}$ describes the
propagation of a photon
that is being
observed at x, propagating along ${\bar {f}},$ and that will be absorbed by J
at $y_{2}.$
$J(y_{2})$ is its
sink.
\end{enumerate}
$y_{1}$ and $y_{2}$ are the intersections of $J(y),$ the charge causality-line
with the x-
causality-cone.
The f's of these two kinds of solutions are related as f and ${\bar{f}}$, as
defined in section II.
This process of creation and annihilation of particles in classical physics
replaces the usual
scheme of retarded and advanced solutions. All solutions are retarded and there
is no
causality violation.
\item The set of states, in the Quantum Mechanics language, labelled by f and
represented
also by (\ref{expqgreen}), in contradistinction to (\ref{cG}), constitutes a
vector space; it
obeys the Principle of Linear Superposition and is, therefore, appropriate for
a quantum
theory representation. This includes the additional hypothesis, required to fit
the
observational data, of adding the amplitudes of probability (not the
probabilities), which is
possible only in a vector space;
\item Considering the finite and macroscopic resolution of our measuring
apparatus, and that
the f-causality lines are unidimensional geometric objects, then, any
determination of, let's
say, a photon $A_{f},$ is unavoidably made  within an uncertainty window,
$\Delta f,$ that
contains the uncertainties in the photon position and  velocity. This lays a
bridge to the
Uncertainty Principle of Quantum Mechanics$^{(2,6)}$
\item A fundamental hypothesis in the derivation of (\ref{expqgreen}) is that
of a constant f.
But f is also the velocity of the propagating fields. Therefore, they freely
propagate along a
causality-cone generator and their interactions are discrete and localized at
points, the
causality-cone vertices. As explained in section II, a causality-cone vertex
corresponds to an
interaction Lagrangian term, like ${\cal
L}_{I}=e{\bar\Psi}_{f}A_{f'}\Psi_{f''}.$ The localized
and discrete radiation  emission at a single point is the most essential and
definitive quantum
aspect of a field theory.  This is a classical formalism with respect to
dealing with classical
fields instead of with operators, whereas it is already a quantum formalism
with respect to
interactions and field propagation.
\item  It is interesting that (\ref{expqgreen}) prohibits self interactions and
vacuum
fluctuations: its delta function requires that each field follow and remain
strictly on their
respective causality-lines, which never cross each other again. For example, an
electron and
its emitted photon follow different causality-lines, with
respectively, $f^2=-1$ and $f^2=0$, and so the probability that this photon
be reabsorbed by this electron is null. On the other hand, as the field
equation have been
changed by the use of directional derivatives (\ref{directional}), this may
compensate for the
absence of these renormalization effects. In other words, one may hope, that
the exact
solutions of the new field equations produce the exact number in effects like
the Lamb shift
and the electron anomalous magnetic moment, where, as well known, these quantum
corrections have so well fitted the data.
\item Although the conformally invariant (\ref{expqgreen}) cannot have any
explicit
dependence on any field mass,  for $\Delta\tau\neq0,$ which is possible only
for non-abelian
fields, it describes the propagation of a massive field. The crux point is
that, in this case,
$f.dx\neq0,$ and so, f does not represent the field velocity anymore. To make a
change for
the physical velocities corresponds to diagonalize the kinetic lagrangian term.
For sets of
non-abelian fields, it can be shown, that their masses are eigenvalues of their
kinetic
Lagrangian term, independently of any interaction, and that they are just
determined by their
correspondingly enlarged spacetime symmetry. This indicates the intrinsically
kinematical
character of the mass. The interested reader can find in the reference [3], in
a compact
communication, an example where, for the $SU(2)\otimes U(1)$ symmetry the weak
boson
mass sector of the Lagrangian of the Weinberg-Salam electroweak theory is
obtained.
\item The geometrical ideas behind (\ref{expqgreen}) is entirely compatible and
consistent
with Lagrangian and variational methods, with a consistent interpretation of
the Noether
theorem and of the Poincar\`e Group algebra$^{(6)}$;
\item From (\ref{expqgreen}) one can develop a consistent, entirely free of
singularity and of
ambiguity, Classical Electrodynamics of a point charge, as done in reference
[7]. In
particular, the Einstein-de Broglie relation $(E,\vec{p})=h\nu(1,\hat{p}),$
that equates the
photon 4-momentum, $p^{\mu}$, to its 4-wave vector times the Planck constant,
is {\it
obtained} from $p^{\mu}=\int dx^{4}\Theta^{\mu\nu}_{f}dS_{\nu},$ if the time
component of f
is taken as a measure of $\nu$, the photon frequency. Then, the Planck constant
can be
identified with a given constant integral over the photon sources and is
closely related to the
fundamental discrete process of radiation emission described in this
f-formalism;
\item These same ideas can be applied to the General Theory of
Relativity$^{(8)}$. For a
spherically symmetric distribution of masses one finds a solution
$g^{\mu\nu}_{f}$, which is
free of singularity and reproduces the Schwarzschild solution, upon following
the recipe
(\ref{classicfield}).
\end{enumerate}

\section{\bf Appendix}

Let us prove (\ref{fec}). Its RHS corresponds to
\begin{equation}
\label{afec}
\lim_{\varepsilon\to0}\int
dx^{4}f^{\beta}g_{\beta\mu}\Biggl\{\nabla_{\nu}\Theta^{\mu\nu}+\Gamma^{\mu}_{\alpha\nu}\Theta^{\alpha\nu}\Biggr\}\theta(\rho-\varepsilon),
\end{equation}
and then,
\begin{equation}
\label{bfec}
\lim_{\varepsilon\to0}\int
dx^{4}\Biggl\{{\biggl(}f^{\beta}g_{\beta\mu}\Gamma^{\mu}_{\alpha\nu}-
\nabla_{\nu}(f^{\beta}g_{\beta\alpha}){\biggr)}\Theta^{\alpha\nu}\theta(\rho-\varepsilon)-
f_{\mu}\Theta^{\mu\nu}\nabla_{\nu}\rho\;\delta(\rho-\varepsilon)\Biggr\}.
\end{equation}
But, $f_{\mu}\Gamma^{\mu}_{\alpha\nu}=\frac{\chi}{2}{\biggl(}{\hbox{\Large
a}}_{f}(V'_{\alpha},f'_{\nu})-({\hbox{\Large a}}_{\alpha},f'_{\nu})
-f.f'({\hbox{\Large
a}}_{\alpha},V'_{\nu}){\biggr)}=\frac{\chi}{2\rho^{2}}{\biggl(}{\hbox{\Large
a}}.R(V'_{\alpha},R'_{\nu})-\rho\;({\hbox{\Large a}}_{\alpha},R'_{\nu})
-R.R'({\hbox{\Large
a}}_{\alpha},V'_{\nu}){\biggr)}$
 and so, from (\ref{t}), the behaviour of the first term of the integrand of
(\ref{bfec}) in the limit
of R$\rightarrow0$ is given by
\begin{equation}
\label{cfec}
\lim_{R\to0}\frac{[R,\rho\;{\hbox{\Large a}}+V(1+{\hbox{\Large
a}}.R)].g.[R,\rho{\hbox{\Large
a}}+V(1+{\hbox{\Large a}}.R)]{\biggl(}{\hbox{\Large
a}}.R(V,R)-\rho\;({\hbox{\Large a}},R)-
R.R'({\hbox{\Large a}},V)\frac{\chi}{2}{\biggr)}}{\rho^{6}}.
\end{equation}
Then, according to the notation used in (\ref{AgBC}),
$$C_{0}={\hbox{\Large a}}.R(V,R)-\rho\;({\hbox{\Large a}},R)-R.R'({\hbox{\Large
a}},V)\frac{\chi}{2}\Longrightarrow C_{2}=-2\chi(\hbox{\Large a},V)+{\cal
O}(R)\rightarrow
c=2$$
$$A_{0}=B_{0}=[R,\rho\;\hbox{\Large a}+V'(1+\rho\hbox{\Large
a}.R)]\Longrightarrow
A_{2}=B_{2}=[\hbox{\Large a},V']+{\cal O}(R)\Longrightarrow a-2=2$$
$$\Longrightarrow p-a=2\Longrightarrow p=6=n\Longrightarrow
N_{6}=\pmatrix{6\cr4}\pmatrix{4\cr2}[\hbox{\Large a},V'].g.[\hbox{\Large
a},V'].{\biggl(}-
2\chi(\hbox{\Large a},V){\biggr)}\equiv0,$$
according to (\ref{NP}) and (\ref{TG}).
Therefore,
\begin{equation}
\label{fGT}
\lim_{\varepsilon\to0}\int
dx^{4}f_{\mu}\Gamma^{\mu}_{\alpha\nu}\Theta^{\alpha\nu}\theta(\rho-\varepsilon)=0
\end{equation}
To find how the second term of the integrand of (\ref{bfec}) behaves in the
limit of
R$\rightarrow0$, we need to find $\nabla_{\nu}f'_{\mu}$ from
$f^{\mu}=\frac{R^{\mu}}{\Delta\tau}$ and $\rho=\Delta\tau.$ The difference
between the
derivatives of $\rho$ and of $\Delta\tau$ tends to zero in the limit of
$\rho\rightarrow0.$ So, it
is irrelevant if we use one or the other in the definition of f; we use the
simplest one,
$\Delta\tau,$ and $\frac{\partial\Delta\tau}{\partial\tau}= -1$ to find:
$$\nabla_{\nu}(g_{\mu\alpha}f^{\alpha})=\frac{1}{\Delta\tau}{\Big\lbrace}\Delta\tau\chi
f'_{\nu}(V'_{\mu}{\hbox{\Large a}}_{f}-{\hbox{\Large
a}}_{\mu})+g_{\mu\nu}+f'_{\nu}(V'-
f')_{\mu}){\Big\rbrace}=$$
$$=\frac{1}{(\Delta\tau)^{3}}{\Big\lbrace}(\Delta\tau)^{2}(g_{\mu\nu}-\chi{\hbox{\Large
a}}_{\mu}R'_{\nu})+\Delta\tau {\biggl(}V'_{\mu}(1+\chi{\hbox{\Large
a}}.R){\biggr)}R'_{\nu}-
R'_{\mu}R'_{\nu}{\Big\rbrace}$$
We just have to replace, in the previous case ,its $C_{0}$ by
$$C_{0}=(\Delta\tau)^{2}(g-\chi{\hbox{\Large a}}R')+\Delta\tau
R'V'(1+\chi{\hbox{\Large a}})-
R'R'$$
Thus, $$C_{2}=2g+{\cal O}(R)\Longrightarrow c=2\Longrightarrow p=6<n=7$$
According to (\ref{NR}), this would produce a divergent result if $N_{6}\neq0,$
but $\Theta$
and its limit are traceless $\Theta^{\mu\nu}g_{\mu\nu}=0$, and so,
$N_{6}=0.$
Therefore, the indeterminacy, $0/0$ remains and a new application of the
L'Hospital rule is
demanded. Then, from (\ref{NP}), for $p=6,\;a=4,\;c=2$, we have
$$N_{6}=\pmatrix{6\cr 2\cr}\pmatrix{2\cr 2\cr}A_{2}.\;g_{0}\;.B_{2}C_{2}=0,$$
and then
$$N_{7}=\dot{N}_{6}=\pmatrix{6\cr 2\cr}\pmatrix{2\cr
2\cr}{\Big\lbrace}A_{3}.\;g_{0}\;.A_{2}C_{2}+A_{2}.\;g_{0}\;.A_{2}C_{2}+A_{2}.\;g_{1}\;.A_{2}C
_{2}+A_{2}.\;g_{0}\;.A_{3}C_{2}+A_{2}.\;g_{0}\;.A_{2}C_{3}{\Big\rbrace}.$$
The terms containing $C_{2}$ will still give a null contribution ($\Theta$ is
traceless).
With
$C_{3}=3(V,\hbox{\Large a})(\chi+\frac{1`}{2})+{\cal O}(R),$
\begin{equation}
\label{dfT3}
{\Big\lbrace}A_{2}.\;g_{0}\;.A_{2}C_{3}{\Big\rbrace}={\Big\lbrace}[V,\hbox{\Large
a}].g.[V,\hbox{\Large a}]gg(9V\hbox{\Large a}+6\hbox{\Large
a}V){\Big\rbrace}\equiv0,
\end{equation}
and then,
\begin{equation}
\label{dfT}
\lim_{\varepsilon\to0}\int
dx^{4}\nabla_{\nu}f_{\alpha}\Theta^{\alpha\nu}\theta(\rho-
\varepsilon)=0.
\end{equation}
The last term in the integrand of (\ref{bfec}) is related to
(\ref{P}-\ref{P0}). So, we can write
\begin{equation}
\label{fP2}
<f.P_{2}>:=\lim_{\varepsilon\to0}\int
dx^{4}f^{\alpha}g_{\alpha\mu}\Theta_{4}^{\mu\nu}(f'_{\nu}-V'_{\nu})\delta(\rho-\varepsilon),
\end{equation}
which is null because
$$\lim_{\rho\to0}\frac{R.\lbrace[R,V'].g.[R,V']\rbrace.g.(R-V\rho)}{\rho^{6}}=0;$$
and
\begin{equation}
\label{fP1a}
<f.P_{1}>:=\lim_{\varepsilon\to0}\int
dx^{4}f^{\alpha}g_{\alpha\mu}{\Big\lbrace}\Theta_{3}^{\mu\nu}(f'-
V')_{\nu})+\Theta_{4}^{\mu\nu}f'_{\nu}\rho\;{\hbox{\Large
a}}_f{\Big\rbrace}\delta(\rho-
\varepsilon),
\end{equation}
which is also null because
$$
\lim_{\rho\to0}\frac{R.\lbrace[R,V'].g.[R,\rho\;\hbox{\Large a}+V'\hbox{\Large
a}.R].(R-
\rho\;V)\rbrace}{\rho^{5}}=
\lim_{\rho\to0}\frac{R.g.\lbrace{[R,V'].g.[R,V']\rbrace}.R\hbox{\Large
a}.R}{\rho^{5}}=0,$$
as they can be easily verified. Finally,
\begin{equation}
\label{fP0a}
<f.P_{0}>:=\int
dx^{4}f^{\alpha}g_{\alpha\mu}{\Big\lbrace}\Theta_{2}^{\mu\nu}(f'-
V')_{\nu})+\Theta_{3}^{\mu\nu}f'_{\nu}\rho\;{\hbox{\Large
a}}_f{\Big\rbrace}\delta(\rho-
\varepsilon)=0,
\end{equation}
as a consequence of (\ref{C}).

\newpage
\begin{center}
FIGURE CAPTION
\end{center}
Creation (${\bar{f}})$ and destruction (f) of particles.\\ f and ${\bar f}$
represent two distinct
types of propagating solutions of $\Box A_{f}=J:$
\begin{enumerate}
\item Along ${\bar f}$ propagates a photon that was emitted at $y_{1}$ and that
has been observed at x.  $\Delta x^{0}=x^{0}-y_{1}^{0}>0.\;$  $J(y_{1})$ is its
source.
\item along f propagates a photon that has been observed at x and that will be
absorbed by J at $y_{2}$. $\Delta x^{0}=x^{0}-y_{2}^{0}<0.$ $J(y_{2})$ is its
sink.
\end{enumerate}
There is no advanced field, only retarded fields. $\Delta x^{0}>0$ and $\Delta
x^{0}<)$ describe creation and destruction of fields, respectively.
\end{document}